\documentclass[12pt]{article}
\usepackage{graphicx}
\usepackage{amsmath}
\usepackage{amsbsy}
\usepackage{amsfonts}
\usepackage{amsthm}
\begin{document}

\title{EYM equations in the presence of q-stars}

\author{Athanasios Prikas}

\date{}

\maketitle

Physics Department, National Technical University, Zografou
Campus, 157 80 Athens, Greece.\footnote{e-mail:
aprikas@central.ntua.gr}

\begin{abstract}
We study Einstein-Yang-Mills equations in the presence of
gravitating non-topological soliton field configurations, of
q-ball type. We produce numerical solutions, stable with respect
to gravitational collapse and to fission into free particles, and
we study the effect of the field strength and the eigen-frequency
to the soliton parameters. We also investigate the formation of
such soliton stars when the spacetime is asymptotically anti de
Sitter.
\end{abstract}

PACS number(s): 04.70.Bw, 04.20.Jb

\newpage

\section{Introduction}

Coupled Einstein-Yang-Mills (EYM) equations have been studied in
various contexts, \cite{EYMeqs-review}. Bartnik and McKinnon
found particle-like non-abelian solutions of the coupled EYM
theory, \cite{Bartnik-McKinnon1,Bartnik-McKinnon2}. The coupling
of the EYM system with a scalar field leads to several theories.
We mention gravitating Skyrmions, \cite{EYMskyrmions}, black hole
solutions in dilaton, \cite{EYMdilaton1,EYMdilaton2}, and massive
dilaton and axion gravity, \cite{EYMmassivedilaton}, and other
field configurations in the EYM-Higgs theory with a Higgs
doublet, \cite{EYMhiggs-doublet} or with a Higgs triplet,
\cite{EYMhiggs-triplet1,EYMhiggs-triplet2,EYMhiggs-triplet3}.

Non-topological solitons appeared initially as an attempt to
explain the structure of hadrons with various ``bag" models, or
independently, as mathematical objects,
\cite{non-top-sol-review}. There are two main kinds of
non-topological solitons and soliton-stars, q-solitons (q-balls
and q-stars) and ``large" non-topological solitons. ``Large"
non-topological solitons are characterized by a potential of the
form $m^2|\Phi|^2(1-|\Phi|^2/|\Phi_0|^2)^2$. Inside the soliton
or the soliton-star, $\phi\simeq\phi_0$,
\cite{non-top-sol-review,large-soliton-stars1,large-soliton-stars2,large-soliton-stars3}
and the kinetic terms arising from the temporal variation of the
scalar field dominate. In q-balls the total energy is minimized
at the minimum of the $\sqrt{U/|\Phi|^2}$,
\cite{qballs-initial,qballs-nonabelian}. Their relativistic
generalizations, may consist of one or two scalar fields,
\cite{qstars-global}, in a Lagrangian with a global $U(1)$
symmetry, or of a non-abelian scalar field in the adjoint
representation of $SU(3)$, \cite{qstars-nonabelian}, or with a
scalar and a fermion field, \cite{qstars-fermion} in
asymptotically flat or anti de Sitter spacetime,
\cite{qstars-AdS}. Q-solitons with local symmetries have also
been investigated. There are charged q-balls,
\cite{qballs-charged}, and charged q-stars, \cite{qstars-charged}.

The purpose of the present work is to find numerical solutions of
the EYM equations in the presence of a Higgs doublet in the
fundamental representation of $SU(2)$, reducing the EYM-Higgs
equations to a field configuration corresponding to a (charged)
q-star. In the absence of the gauge field, the equations of
motion give rise to a gravitating non-topological soliton, when
using a special potential for which
$\omega_E\equiv\sqrt{U/|\Phi|^2}_{\textrm{min}}<m$ where $m$ is
the mass of the free particles and imposing an harmonic
time-dependence on the scalar field with the frequency equal to
$\omega_E$. Our gravitating soliton is \emph{non-topological} in
the sense that $\Phi,U\rightarrow0$ for $\rho\rightarrow\infty$
according to \cite{non-top-sol-review}. It is a \emph{q-type}
non-topological soliton in the sense that in the absence of both
gravitational and gauge fields one can find by simple
calculations that this spherically symmetric Higgs field rotates
within its symmetry space with a frequency $\omega_E$ equal to
the minimum of the $\sqrt{U/|\Phi|^2}$ quantity. The difference
between this soliton and the usual non-abelian q-balls is that the
symmetry space in the case of non-abelian q-balls is the entire
$SU(3)$ space but in our case is an abelian $U(1)$ subgroup of the
$SU(2)$, though both field configurations are non-abelian, and
that non-abelian q-balls consist of fields in the adjoint
representation of $SU(3)$, when our Higgs are in the fundamental
representation. We also study the solutions of the above soliton
when the spacetime is asymptotically anti de Sitter. In any case,
we find an analytical solution for the scalar field within the
soliton, using the approximation known by the study of q-stars.

\section{EYM equations in the framework of soliton stars}

The action of a Yang-Mills gauge field coupled to a Higgs scalar
in the fundamental representation and to gravity in $3+1$
dimensions is:
\begin{equation}\label{1}
S_{\textrm{HEYM}}=\int\left(-\frac{R-2\Lambda}{16\pi
G}+\frac{1}{4K\textrm{g}^2}\textrm{Tr}F_{\mu\nu}F^{\mu\nu}+
(D_{\mu}\Phi)^{\dagger}(D^{\mu}\Phi)- U\right)\sqrt{-g}d^4x\ ,
\end{equation}
where $\Lambda$ stands for the cosmological constant, $U$ is the
potential and:
\begin{equation}\label{2}
\begin{split}
D_{\mu}\Phi&=\partial_{\mu}\Phi-\imath A_{\mu}\Phi \\
F_{\mu\nu}&=\partial_{\mu}A_{\nu}-\partial_{\nu}A_{\mu}-\imath[A_{\mu},A_{\nu}]\
.
\end{split}
\end{equation}
One may use the
$F_{\mu\nu}=\partial_{\mu}A_{\nu}-\partial_{\nu}A_{\mu}-\imath\textrm{g}[A_{\mu},A_{\nu}]$
form and eq. \ref{2} may be reproduced by redefining
$A_{\mu}\rightarrow A_{\mu}/\textrm{g}$, where $\textrm{g}$ is
the field strength, or coupling constant. The one-form gauge field
$A$ is: $A\equiv
A_{\mu}dx^{\mu}\equiv\mathbf{T}_aA^a_{\mu}dx^{\mu}$, with
$\mathbf{T}_a=\frac{1}{2}\tau_a$ and $\tau_a$ the Pauli matrices.
The factor $K$ appearing in the action is defined by the relation
$\textrm{Tr}(\mathbf{T}_a\mathbf{T}_b)=K\delta_{ab}$, here we find
$K=1/2$. We will choose a general spherically symmetric field
configuration, defining:
$n^a\equiv(\sin\vartheta\cos\varphi,\sin\vartheta\sin\varphi,\cos\vartheta)$
and:
\begin{equation}\label{3}
\mathbf{T}_{\rho}=n^a\mathbf{T}_a\
,\hspace{1em}\mathbf{T}_{\vartheta}=\partial_{\vartheta}\mathbf{T}_{\rho}
\,\hspace{1em}\mathbf{T}_{\varphi}=\frac{1}{\sin\vartheta}\partial_{\varphi}\mathbf{T}_{\rho}\
.
\end{equation}
The gauge and scalar fields take respectively the forms:
\begin{align}\label{4}
&A=a\mathbf{T}_{\rho}+\imath(1-\textrm{Re}\omega)[\mathbf{T}_{\rho},d\mathbf{T}_{\rho}]+
\textrm{Im}\omega\mathbf{T}_{\rho}= \nonumber\\
&a\mathbf{T}_{\rho}+[\textrm{Im}\omega\mathbf{T}_{\vartheta}+(\textrm{Re}\omega-1)
\mathbf{T}_{\varphi}]d\vartheta+[\textrm{Im}\omega\mathbf{T}_{\varphi}+
(1-\textrm{Re}\omega)\mathbf{T}_{\vartheta}]\sin\vartheta d\varphi
,
\end{align}
\begin{equation}\label{5}
\Phi=\sigma\exp (\imath\xi\mathbf{T}_{\rho})|b\rangle
\end{equation}
with $\sigma=\sigma(\rho,t)$, $\xi=\xi(\rho,t)$, $|b\rangle$ a
constant unit vector of the internal $SU(2)$ space of the scalar
(Higgs) field and $a=a_0dt+a_{\rho}d\rho$.

Our purpose is to find a special spherically symmetric, static
field configuration resulting to a non-topological soliton. For
the sake of simplicity we choose $a_{\rho}=0$ and $a_0=a_0(\rho)$,
$\sigma(\rho,t)=\sigma(\rho)$ and $\xi=\omega_Et$ in order to
form a static configuration. The ansatz
$\sigma(\rho,t)=\sigma(r)$ and $\xi=\omega_Et$ is the obvious
generalization to the $\phi(\rho,t)=\sigma(\rho)e^{\imath\omega
t}$ ansatz, known from q-solitons. The spherically symmetric,
static metric is written:
\begin{equation}\label{6}
ds^2=-\frac{1}{B}dt^2+\frac{1}{A}d\rho^2+\rho^2d\vartheta^2+\rho^2\sin^2\vartheta
d\varphi^2\ .
\end{equation}
Inserting eqs. \ref{4}-\ref{6} in eq. \ref{1} we find for the
matter action:
\begin{align}\label{7}
S_{\textrm{matter}}=\int\frac{\rho^2\sin\vartheta}{\sqrt{AB}}\left[-\frac{1}{2\textrm{g}^2}
\left(a_0'^2AB+2\frac{|\omega|^2a_0^2}{\rho^2}+\frac{(|\omega|^2-1)^2}{\rho^4}\right)
+\sigma'^2A \right. \nonumber\\ \left.
-\frac{1}{4}(\omega_E-a_0)^2\sigma^2B+\frac{\sigma^2}{2\rho^2}
[(\textrm{Re}\omega-\cos(\omega_Et))^2+(\textrm{Im}\omega-\sin(\omega_Et))^2]-U
\right]
\end{align}
In order the action to be time-independent we may choose
$\textrm{Re}\omega=\cos(\omega_Et)$ and
$\textrm{Im}\omega=\sin(\omega_Et)$, but this choice is not a
solution of the equation of motion for $\omega$, or $\omega=0$,
which is a solution of the equation of motion, so, our solution is
embedded abelian.

In the absence of gauge fields, the Higgs field forms a soliton,
which when coupled to gravity is a soliton star. This simplifies
drastically the equations of motion and separate the total
$3$-dimensional space in three regions, the interior, the very
thin surface and the exterior, where the matter scalar can be
regarded as zero. Let $U$ be equal to:
\begin{equation}\label{8}
U=m^2\sigma^2\left(1-\frac{4\sigma^2}{m^2}+\frac{16}{3m^4}\sigma^4\right)\
.
\end{equation}
The quantity $\sqrt{U/\sigma_0^2}$ is smaller than the mass of
the free particles, so for the ``q-ball-type solution", we write
$\sigma=\sigma_0$, where $\sigma_0$ is the value that minimizes
the above quantity, and
$\omega_E=\sqrt{U/\sigma_0^2}_{\textrm{min}}$ is the minimum
energy frequency, \cite{qballs-initial,qballs-nonabelian}. In the
original papers referring to q-solitons, the non-abelian scalar
field is in the adjoint representation of the $SU(2)$. The
results can be straightforward generalized for the fundamental
representation. As one can see, $\sigma_0$ and $\omega_E$ are of
the same order of magnitude for q-type solitons (q-balls and
q-stars), in contrast with the non-topological soliton (stars),
investigated in \cite{non-top-sol-review}, where $\omega_E\ll
|\phi|,m$. We define:
\begin{equation}\label{9}
W=\left(\frac{d\Phi}{dt}\right)^{\dagger}\left(\frac{d\Phi}{dt}\right)\
,\hspace{1em}V=\left(\frac{d\Phi}{d\rho}\right)^{\dagger}\left(\frac{d\Phi}{d\rho}\right)\
.
\end{equation}
The ``q-ball-type solution" (in the so-called ``thick-wall"
approximation) means that the scalar field within the soliton is
approximately constant, $W$ and $U$ are of the same order of
magnitude, when $V$ is negligible and the energy density is $\sim
m^4$. Gravity becomes important when $R\sim 8\pi GM_R$ where $M_R$
is the mass trapped within a sphere of radius $R$. We find that
$R\sim 1/(8\pi Gm^4)^{1/2}$. We define:
\begin{equation}\label{10}
\epsilon\equiv\sqrt{8\pi Gm^2}\ .
\end{equation}
If $\sigma(0)\simeq\sigma_0$ and $\sigma(\rho)\simeq0$ for
$\rho>R$, then $V\sim\epsilon^2m^4$. For $m\sim GeV$ the
$O(\epsilon)$ quantities are negligible. The equation of motion
for the Higgs field is:
\begin{equation}\label{11}
A\left[\frac{d^2\sigma}{d\rho^2}+\frac{1}{2}\left(\frac{4}{\rho}+\frac{1}{A}
\frac{dA}{d\rho}-\frac{1}{B}\frac{dB}{d\rho}\right)\frac{d\sigma}{d\rho}\right]+
\frac{B\theta_0^2 \sigma}{4}-\frac{dU}{d\sigma^2}\sigma=0\ ,
\end{equation}
where we define a useful quantity:
\begin{equation}\label{12}
\theta_0=\omega_E-a_0\ .
\end{equation}

We make the following rescalings:
\begin{align}\label{13}
&\tilde{\rho}=2m\rho \ ,\hspace{1em}
\tilde{\omega}_E=\frac{\omega_E}{2m} \ ,\hspace{1em}
\tilde{a}_0=\frac{a_0}{2m} \ ,\hspace{1em}
\tilde{\sigma}=\frac{\sigma}{\frac{m}{2}}\ ,\nonumber\\
&\tilde{r}=\epsilon\tilde{\rho}\
,\hspace{1em}\tilde{\textrm{g}}=\textrm{g}\epsilon^{-1}\
,\hspace{1em}\widetilde{\Lambda}\equiv\frac{\Lambda}{8\pi Gm^4}\ .
\end{align}
With the above rescalings $\tilde{r}\sim1$. In our solutions we
find $\widetilde{\Lambda}\sim1$, so, the ``true" cosmological
constant, $\Lambda$, is of $O(8\pi Gm^4)$ order. From now on the
prime denotes the differentiation with respect to $\tilde{r}$. The
potential takes a simple form, regarding $m=1$, namely:
$U=1/4\sigma^2(1-\sigma^2+1/3\sigma^4)$. This potential has
minimum at $\sigma_0^2=1.5$, which gives $\omega_E=0.5$. Dropping
from the Lagrange equation the tildes and the $O(\epsilon)$
quantities, we find an analytical solution for the scalar field:
\begin{equation}\label{14}
\sigma^2=1+\theta_0B^{1/2}\
,\hspace{1em}U=\frac{1}{12}(1+\theta_0^3B^{3/2})\
,\hspace{1em}W=\theta_0^2B(1+\theta_0B^{1/2})\
,\hspace{1em}V\simeq0\ .
\end{equation}
The above equation holds true within the soliton interior. The
surface width is of $m^{-1}$. The scalar field within the surface
changes rapidly from a $\sigma\simeq\sigma_0$ value at the inner
edge of the surface to a $\sigma\simeq0$ value at the outer one.
Integrating the equation of motion within the surface we find:
\begin{equation}\label{15}
V+W-U=0\ .
\end{equation}
At the inner edge of the surface $\sigma'=0$ so as to match the
interior with the surface solution. Using eqs. \ref{14}, \ref{15}
we find the eigenvalue equation for the new field $\theta_0$:
\begin{equation}\label{16}
{\theta_0}_{\textrm{sur}}=\frac{A^{1/2}_{\textrm{sur}}}{2}=\frac{B_{\textrm{sur}}^{-1/2}}{2}\
,
\end{equation}
where ${\theta_0}_{\textrm{sur}}$ is the value of $\theta_0$
within the thin surface.

The Einstein equations are: $G_{\mu}^{\ \nu}=8\pi GT_{\mu}^{\
\nu}-\Lambda\delta_{\mu}^{\ \nu}$, where the energy-momentum
tensor, $T_{\mu\nu}$, is defined by the relation:
\begin{align}\label{17}
T_{\mu\nu}=\frac{2}{\textrm{g}^2}\textrm{Tr}\left(g^{\alpha\beta}F_{\mu\alpha}F_{\nu\beta}
-\frac{1}{4}g_{\mu\nu}F^{\alpha\beta}F_{\alpha\beta}\right)+
\nonumber\\ (D_{\mu}\Phi)^{\dagger}(D_{\nu}
\Phi)+(D_{\mu}\Phi)^T(D_{\nu}\Phi)^{\ast}-g_{\mu\nu}[g^{\alpha\beta}(D_{\alpha}\Phi)^{\dagger}
(D_{\beta}\Phi)]-g_{\mu\nu}U\ .
\end{align}
Dropping the $O(\epsilon)$ quantities, the independent Einstein
equations take the simple form:
\begin{equation}\label{18}
\frac{A-1}{r^2}+\frac{A'}{r}=-U-W-\frac{\theta_0'^2}{2\textrm{g}^2}AB-\Lambda\
,
\end{equation}
\begin{equation}\label{19}
\frac{A-1}{r^2}-\frac{A}{B}\frac{B'}{r}=W-U-\frac{\theta_0'^2}{2\textrm{g}^2}AB-\Lambda\
,
\end{equation}
and the equation of motion for the gauge field is:
\begin{equation}\label{20}
\theta_0''+\left(\frac{2}{r}+\frac{A'}{2A}+\frac{B'}{2B}\right)\theta_0'-\frac{\textrm{g}^2
\theta_0(1+\theta_0B^{1/2})}{2A}=0\ .
\end{equation}

\begin{figure}
\centering
\includegraphics{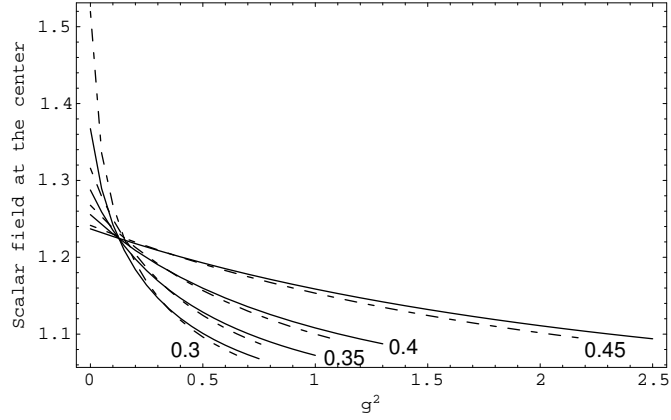}
\caption{The value of $\sigma(0)$ as a function of the coupling
constant $g^2$. Dashed lines correspond to $\Lambda=-0.02$ and
solid lines to asymptotically flat spacetime.} \label{figure1}
\end{figure}

\begin{figure}
\centering
\includegraphics{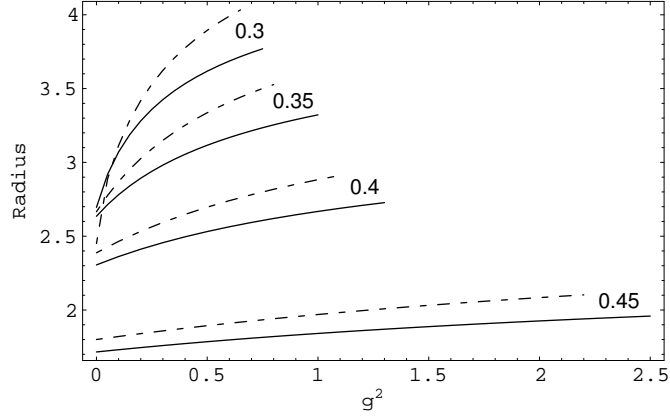}
\caption{The radius of the soliton as a function of $g^2$.}
\label{figure2}
\end{figure}

\begin{figure}
\centering
\includegraphics{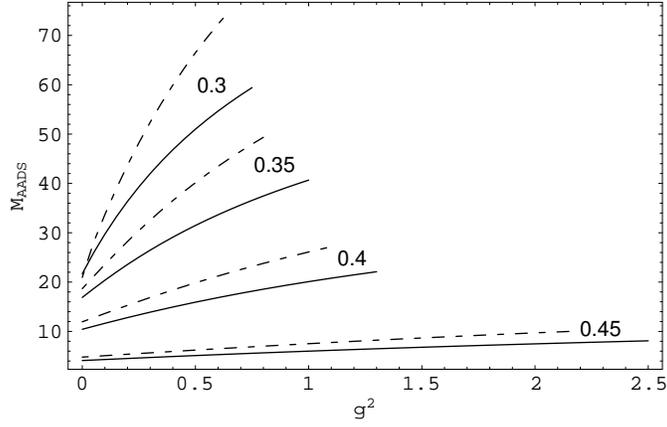}
\caption{The asymptotically anti de Sitter mass,
$M_{\textrm{AADS}}$, as a function of $g^2$.} \label{figure3}
\end{figure}

\begin{figure}
\centering
\includegraphics{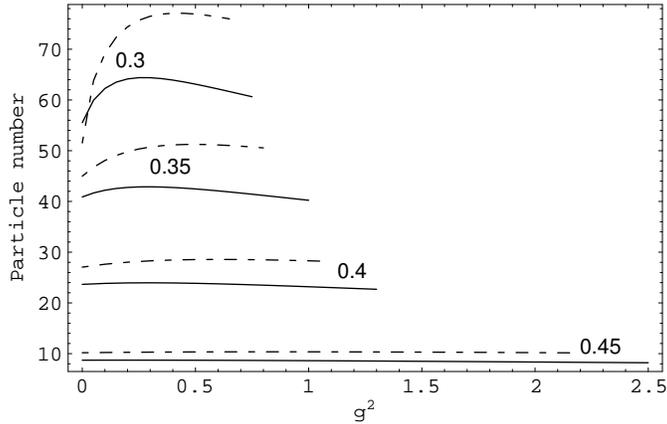}
\caption{The particle number, $N$, of the soliton as a function
of $g^2$.} \label{figure4}
\end{figure}

\begin{figure}
\centering
\includegraphics{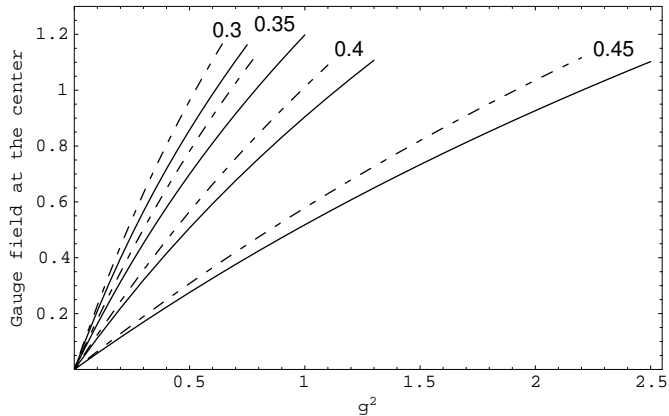}
\caption{The value of the gauge field, $\alpha$, at the center of
the soliton as a function of $g^2$.} \label{figure5}
\end{figure}

\begin{figure}
\centering
\includegraphics{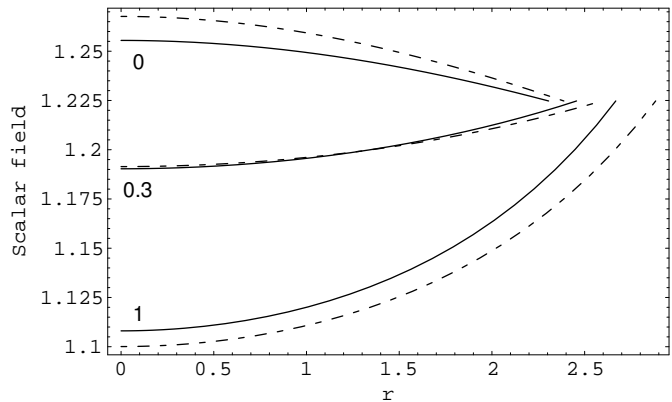}
\caption{$\sigma$ as a function of $r$.} \label{figure6}
\end{figure}

There are three Noether currents, related to the generators of
the Lie algebra, which are given by the relation:
\begin{equation}\label{21}
j_{0\alpha}=\left(\begin{array}{cc}
  \frac{\partial L}{\partial(\partial_0\Phi)} & \frac{\partial
  L}{\partial(\partial_0\Phi^{\ast})}
\end{array}\right)\left(\begin{array}{cc}
  \imath T_{\alpha} & 0 \\
  0 & -\imath T_{\alpha}
\end{array}\right)\left(\begin{array}{c}
  \Phi\ \\
  \Phi^{\ast}
\end{array}\right)\ .
\end{equation}
We can find that:
\begin{equation}\label{22}
\begin{split}
j_{01}&=\frac{1}{2}\sigma^2\theta_0\sin\vartheta\cos\varphi \\
j_{02}&=\frac{1}{2}\sigma^2\theta_0\sin\vartheta\sin\varphi \\
j_{03}&=\frac{1}{2}\sigma^2\theta_0\cos\vartheta\
\end{split}
\end{equation}
and
\begin{equation}\label{23}
j_0\equiv\sqrt{j_{01}^2+j_{02}^2+j_{03}^2}=\frac{1}{2}\sigma^2\theta_0\
.
\end{equation}
The particle number is:
\begin{equation}\label{24}
N=2\pi\int\sigma^2\theta_0\sqrt{\frac{A}{B}}r^2dr\ .
\end{equation}
The asymptotically anti de Sitter energy of the field
configuration, $M_{\textrm{AADS}}$, dropping the $O(\epsilon)$
quantities, may be calculated by $-T_0^{\ 0}$, or by:
\begin{equation}\label{25}
M_{\textrm{AADS}}=4\pi r\left(1-A(r)-\frac{1}{3}\Lambda
r^2+\frac{\textrm{g}^2N^2}{32\pi^2r^2}\right)\ ,\hspace{1em}
r\rightarrow\infty\ ,
\end{equation}
resulting form the (unrescaled) relation:
$$A(\rho)=1-\frac{2GM_{\textrm{AADS}}}{\rho}-
\frac{1}{3}\Lambda\rho^2+\frac{G\textrm{g}^2N^2}{4\pi\rho^2}\ ,
\hspace{1em} \rho\rightarrow\infty\ ,$$ with the appropriate
rescalings. We numerically solve the coupled system of eqs.
\ref{18}-\ref{20}. Eqs. \ref{14}, \ref{24} and \ref{25} give some
of the parameters of the soliton.

We will now discuss the sufficient conditions for the existence
of q-solitons in theories with local symmetries,
\cite{sufficient}, and prove that our soliton meets them. These
conditions refer to solitons with zero overall non-abelian
charge, which is the case for the embedded abelian solutions. In
this case, let $\hat{B}$ be the generator of a global $U(1)$
symmetry (our theory possesses such a symmetry) and
$\hat{T}^{\alpha}$ the generators of the non-abelian group. The
energy can be written as, \cite{sufficient}:
\begin{align}\label{sufficient1}
\mathcal{E}_{\lambda,\xi}=\int\sqrt{-g}
d^3x(p^{\dagger}p+|\partial_i\phi|^2+U)\nonumber\\
-\lambda\left[\int\sqrt{-g} d^3x\hat{B}-Q\right]-\int\sqrt{-g}
d^3x\xi^{\alpha}(x)\hat{T}^{\alpha}\ ,
\end{align}
with
$$p(x)=-\imath[\lambda\hat{B}+\xi^{\alpha}(x)\hat{T}^{\alpha}]\phi\
.$$ The equations for the Lagrange multipliers, $\lambda$ and
$\xi^{\alpha}$ are:
\begin{equation}\label{sufficient2}
\lambda\phi^{\dagger}\hat{B}\hat{T}^{\alpha}\phi+\xi^{\beta}\phi^{\dagger}\{\hat{T}^{\alpha}
,\hat{T}^{\beta}\}\phi=0\ ,\hspace{1em} a=1,...,\textrm{dim}(G)\ ,
\end{equation}
\begin{equation}\label{sufficient3}
\int\sqrt{-g}
d^3x[\lambda\phi^{\dagger}\hat{B}^2\phi+\xi^{\beta}(x)\phi^{\dagger}\hat{B}\hat{T}^{\beta}
\phi]=Q\ .
\end{equation}
A q-soliton exists if the system of equations
\ref{sufficient2}-\ref{sufficient3} has a solution. As one can
easily see in the thin-wall limit we discuss in the present work,
the \ref{sufficient2}-\ref{sufficient3} become a system of linear
equations in $\lambda$ and $\xi$ and can easily be solved. There
is also another condition for the existence of q-solitons: the
total energy energy should be less than the energy of the free
particles with the same charge. This condition can be handled
only numerically: We solve the coupled system of eqs.
\ref{18}-\ref{20} and include the solution in our figures only
when the soliton is stable with respect to fission into free
particles.

\section{Conclusions}

We studied a spherically symmetric doublet of scalar Higgs fields
in the fundamental representation of the $SU(2)$ group, coupled to
a gauge field. The EYM-Higgs equations reduced to an
easy-manageable system of equations, corresponding to (charged)
soliton-stars, by choosing a special potential for the Higgs field
and imposing a certain time-dependence to it. The independent
parameters are the coupling constant $\textrm{g}$, which
represents the ``strength" of the gauge field, the
${\theta_0}_{\textrm{sur}}$ eigenvalue, straightforward connected
to the surface gravity through eq. \ref{16} and equal to the
eigen-frequency of the soliton star in the absence of gauge
fields, and the cosmological constant. All the matter and metric
field configurations correspond to \emph{stars} and not to black
holes (i.e.: no horizon and no anomalies at the center are
formed), they are time-independent and stable with respect to
gravitational collapse. All the field configurations discussed in
the above section are stable with respect to decay into free
particles, because their asymptotically anti de Sitter mass is
smaller than the energy of the free particles. When the energy of
the free fields with the same particle number, equal to $mN$ with
$m$ unity for the sake of convenience, tends to become larger than
the soliton energy, we interrupt calculations. The field
configuration in the absence of gravitational and gauge fields
corresponds to a soliton with
$\omega_E=\sqrt{U/\sigma^2}|_{\textrm{min}}$,
$W=\omega_E^2\sigma^2/4=\omega_E^2(1+\omega_E)/4$,
$U=(1+\omega_E^3)/12$ and $V\simeq0$ as one can find by some
simple algebra, i.e.: corresponds to a q-ball type soliton. In the
absence of gauge fields, but in the presence of gravity the field
configuration corresponds to a soliton star with
$\omega_EB=\sqrt{U/\sigma^2}|_{\textrm{min}}$,
$W=B\omega_E^2(1+\omega_EB^{1/2})/4$, $U=(1+\omega_E^3B^{3/2})/12$
and $V\simeq0$ which corresponds to a q-type soliton star.

The numbers within the figures \ref{figure1}-\ref{figure5} denote
the value of ${\theta_0}_{\textrm{sur}}$, which plays the same
role as the frequency in solitons with global symmetries and the
numbers within the figure \ref{figure6} denote $g^2$. All the
field configurations depicted in our figures are stable with
respect to fission into free particles.

The total mass increases with the increase of the coupling
constant due to the electrostatic-type terms arising in the
$-T_0^{\ 0}$ component of the energy-momentum tensor. The value
of $\sigma$ at the center of the soliton decreases due to the
electrostatic repulsion between the different parts of the
soliton. The influence of the cosmological constant to the energy,
particle number and radius of the soliton is clear. The values of
the above parameters are in general larger in an anti de Sitter
spacetime.

\vspace{1em}

\textbf{ACKNOWLEDGEMENTS}

\vspace{1em}

I wish to thank N. D. Tracas and P. Manousselis for helpful
discussions. I also acknowledge partial financial support from the
``Thales-NTUA" programme.

\end{document}